# Spin Relaxation of Conduction Electrons in Semiconductors Due to Interaction with Nuclear Spins

Yuriy V. Pershin and Vladimir Privman

*Center for Quantum Device Technology, Clarkson University,*

*Potsdam, New York 13699-5720, USA*

**Abstract:** Relaxation of conduction electron spins in a semiconductor owing to the hyperfine interaction with spin-½ nuclei, in zero applied magnetic field, is investigated. We calculate the electron spin relaxation time scales, in order to evaluate the importance of this relaxation mechanism. Master equations for the electron spin density matrix are derived and solved. Polarized nuclear spins can be used to polarize the electrons in spintronic devices.

Dynamics of electron spins in semiconductor nanostructures has become of central interest in recent years because of the promise of spintronic devices [1]. In these devices, the information is encoded in the spin state of individual electrons. Operation of a spintronic device requires efficient spin injection into a semiconductor, spin manipulation, control and transport, and also spin detection. One of the first spintronic devices proposed was the Datta-Das field-effect spin-transistor, based on the electrical control of the spin-orbit interaction [2]. Experimental work towards implementations of this proposal has been reported recently [3,4]. A review of and list of references to the latest achievements in spintronics can be found in [5].

Once injected into a semiconductor, electrons experience spin-dependent interactions with the environment, which can cause relaxation. Various relaxation mechanisms of the electron-spin state can result owing to coupling to phonons, magnetic and nonmagnetic impurities, nuclear spins, spins of other electrons [6], etc.; these mechanisms are due to the magnetic and spin-orbit interactions [7,8]. It is important to identify the primary mechanisms of relaxation for a particular system and evaluate how they limit the spin phase coherence.

The main mechanisms of relaxation can be different for bounded and conduction electrons. Several recent works have identified the electron-nuclear spin relaxation mechanism as dominant at low temperatures, for electrons bounded in semiconductor quantum dots [9-13] or at donor impurities [13,14]. Three dominant spin-relaxation mechanisms for conduction electrons were suggested [15-18] and confirmed experimentally (see references in [8]); these are the Elliot-Yafet [15,16], D'yakonov-Perel [17] and Bir-Aronov-Pikus [18] mechanisms. Recent achievements in semiconductor device fabrication of high-purity samples with specified symmetry [19] hold promise of eliminating or at least significantly suppressing these spin relaxation mechanisms. Then, another mechanism of electron spin relaxation, that by nuclear spins, will be dominant for conduction electrons.

Electron-nuclear spin interactions in solids have attracted much interest recently [20-27]. Typically, semiconductor materials contain of at least one stable elemental isotope with nonzero nuclear spin. Interaction of spin-polarized electrons with unpolarized nuclear spins results in the dynamic nuclear spin polarization [20-23] and contributes to the residual conductivity of metals [24]. The level of nuclear spin polarization can be probed by measurements of the resistance [20-23], precession frequency [25], position of the ferromagnetically ordered ground state of the two-dimensional electron gas [26], etc. The interplay between nuclear magnetism and superconductivity was studied in [27] for the first time.

In the present work, we investigate spin relaxation and polarization of conduction electron owing to their interaction with nuclear spins in semiconductor structures in zero applied magnetic field. The electron and nuclear spins are coupled through hyperfine interaction, which is most efficient in zero magnetic field because then the restrictions imposed by the energy conservation, i.e., by the large difference between the electron and nuclear Zeeman splitting, are not present. Moreover, external magnetic field is unlikely to be utilized in spintronic devices. Using density matrix formalism, we obtain Bloch-type equations describing evolution of the electron spin polarization and calculate longitudinal and transverse spin relaxation times. In this article we concentrate on the influence of the nuclear spins on the electron spins, which is reasonable for low electron densities; inclusion of the back-reaction of the electrons on the nuclear polarization is straightforward. For a single electron moving in the background of nuclear spins, the electron spin polarization relaxes with time to the nuclear spin polarization. We find that the transverse relaxation time depends on the degree of the nuclear spin polarization. It is equal to the longitudinal relaxation time if nuclear spins are not polarized, and is twice longer for 100% nuclear spin polarization.

**The Model.** We study conduction electron spin dynamics due to the interaction with nuclear spins at zero magnetic field, within a model based on the density matrix approach for the spin variable, while the spatial motion is considered semi-classically. Indeed, the spin relaxation time is typically orders of magnitude larger than the spatial-motion relaxation. Thus, we assume that the electrons move along trajectories, which are defined by elastic and inelastic scattering events, gradients of potentials, etc., with an average velocity. We are not interested in the exact electron trajectory, and the wave-packet structure of its wave function, since an homogeneous nuclear spin polarization will be assumed. All the information on the spatial electron motion in this semi-classical treatment is lumped in the average electron velocity $v$. The electron Zeeman splitting energy due to the interaction with polarized nuclear spins is rather small; for 100% nuclear spin polarization in GaAs it is of the order of 1K [28]. If the electron temperature is of the same order of magnitude or smaller, then the electron-nuclear spin interaction could strongly affect the electron spatial motion. For example, it was suggested that local nuclear spin polarization can confine the electrons into quantum wires [29]. In the present paper we focus on the opposite case: the effect of the electron-nuclear spin coupling on the electron spatial motion is neglected. This is appropriate, for instance, for spintronic devices operating at room temperature.

The electron interacts with those spin-½ nuclei that are located near its trajectory. Consideration only of nuclei with spin ½ allows derivation of tractable expressions for relaxation times. Relaxation of conduction electron spins by nuclear spins larger than ½, taking into account quadrupole interaction effects, will be considered in a future work. The hyperfine interaction of the electron and nuclear spins is given by the Fermi contact Hamiltonian [30],

$$H = \vec{\sigma} \cdot \sum_i A_i(t) \vec{I}_i , \qquad (1)$$

where $\vec{\sigma}$ is the Pauli-matrix vector corresponding to the electron spin, while $\vec{I}_i$ similarly represents the spin of the $i^{th}$ nucleus, and $A_i$ is the hyperfine coupling constant of the $i^{th}$ nuclear spin to the electron spin, $A_i = \frac{8\pi}{3} g_0 \mu_B \mu_i |\psi(\vec{r}_i,t)|^2$. Here $\mu_B$ is the Bohr magneton, $\mu_i$ is the magnetic moment of the $i^{th}$ nucleus located at position $\vec{r}_i$, and $\psi(\vec{r}_i,t)$ is the time-dependent orbital part of the electron wave function at the nucleus position. In our semi-classical approximation, the electron-nuclear spin hyperfine coupling constant $A_i$ is assumed to be constant, $\overline{A}_i$, during a small time interval from $t_i$ to $t_i + \delta t_i$, and zero otherwise. The time interval $\delta t_i$ is the effective time of the electron-nuclear spin interaction, as the electron's trajectory takes it near the $i^{th}$ nucleus. Moreover, we assume that at any time, the electron either interacts with one nuclear spin or does not interact at all, that is $t_i + \delta t_i < t_{i+1}$.

We will study the evolution of the electron-spin density matrix due to consecutive interactions with nuclear spins along the electron's trajectory. Assuming that between collisions, the spin-up and spin-down electron-spin orientations are degenerate, the relative phases of the density matrix elements are unchanged and it can be taken constant. Let us point out that, in addition to the applied magnetic field being zero, this assumption also implies that other spin-dependent interactions are weak. Since some of these might be needed to control spin-dependent transport in spintronics, notably, the spin-orbit interaction, our calculation applies for motion between the spintronic "gate functions."

Thus, we assume that the electron spin density matrix evolves only by interactions with nuclear spins. Before the interaction of the electron with the $i^{th}$ nuclear spin, at times $t_{i-1} + \delta t_{i-1} < t < t_i$, the electron spin and $i^{th}$ nuclear spin are uncorrelated. The electron spin is described by the density matrix $\rho_e^{i-1} = \begin{pmatrix} \rho_{00}^{i-1} & \rho_{01}^{i-1} \\ \rho_{10}^{i-1} & \rho_{11}^{i-1} \end{pmatrix}$, with $\left(\rho_{10}^{i-1}\right)^* = \rho_{01}^{i-1}$ and real $\rho_{11}^{i-1} = 1 - \rho_{00}^{i-1}$. The following parameterization is used for the $i^{th}$ nuclear spin density matrix,

$$\rho_i = \frac{1}{2}\begin{pmatrix} 1 + P_i^z & P_i^x - iP_i^y \\ P_i^x + iP_i^y & 1 - P_i^z \end{pmatrix}, \qquad (2)$$

where $\vec{P}_i$ is the (real) polarization vector of the $i^{th}$ nuclear spin. At $t = t_i$, the initial density matrix of the two-spin system (the electron and nuclear spins) is given by

$$\rho(t_i) = \rho_e^{i-1} \otimes \rho_i . \qquad (3)$$

The evolution of the two-spin system during the time interval $\delta t_i$, is given by $i\hbar \frac{\partial \rho}{\partial t} = [H, \rho]$. After the interaction, we trace the total density matrix (3) over the nuclear spin. This yields the reduced density matrix of the electron spin at $t_i + \delta t_i$, after the interaction, $\rho_e^i = tr_{I_i} \rho(t_i + \delta t_i)$. It is assumed that electron interacts with each nuclear spin only once. Under these assumptions, we can obtain recurrence equations for the reduced density matrix elements of the electron spin.

Let us consider this calculation in detail. In the two-spin space, the total wave function is $|\varphi\rangle = b_0|\Uparrow\uparrow\rangle + b_1|\Uparrow\downarrow\rangle + b_2|\Downarrow\uparrow\rangle + b_3|\Downarrow\downarrow\rangle$, where $\Uparrow, \Downarrow$ and $\uparrow, \downarrow$ denote electron and nuclear spin states, respectively. We write the wave function as a column vector of $b_{k=0,1,2,3}$. In this representation, the Hamiltonian describing the electron-nuclear spin interaction, cf. (1), takes the form

$$H_i = 2\bar{A}_i \begin{pmatrix} 0 & 0 & 0 & 0 \\ 0 & 0 & 1 & 0 \\ 0 & 1 & 0 & 0 \\ 0 & 0 & 0 & 0 \end{pmatrix} + \bar{A}_i \begin{pmatrix} 1 & 0 & 0 & 0 \\ 0 & -1 & 0 & 0 \\ 0 & 0 & -1 & 0 \\ 0 & 0 & 0 & 1 \end{pmatrix}. \quad (4)$$

The evolution equation yields,

$$\rho(t_i + \delta t_i) = e^{-\frac{i}{\hbar}H_i \delta t_i} \rho(t_i) e^{\frac{i}{\hbar}H_i \delta t_i}. \quad (5)$$

It turns out that all the calculations required to evaluate (5), can be carried out in closed form. After some lengthy algebra, we get the recurrence equations for the elements of the electron spin density matrix,

$$\rho_{00}^i = \rho_{00}^{i-1} \cos^2(2a_i) + \frac{1}{2}(1 + P_i^z)\sin^2(2a_i) + \frac{\sin(4a_i)}{2}\left(P_i^x \operatorname{Im}\rho_{10}^{i-1} - P_i^y \operatorname{Re}\rho_{10}^{i-1}\right), \quad (6)$$

$$\rho_{10}^i = \rho_{10}^{i-1}\left(\cos^2(2a_i) + iP_i^z \frac{\sin(4a_i)}{2}\right) + i(P_i^x + iP_i^y)\sin(2a_i)\left(\frac{e^{-i2a_i}}{2} - \rho_{00}^{i-1}\cos(2a_i)\right), \quad (7)$$

$$\rho_{11}^i = 1 - \rho_{00}^i, \quad \rho_{01}^i = (\rho_{10}^i)^*, \quad (8)$$

where

$$a_i = \frac{\bar{A}_i \delta t_i}{\hbar}. \quad (9)$$

It is convenient to rewrite Eqs. (6)-(8) in terms of the electron spin polarization vector, defined similarly to the nuclear spin polarization vector (see Eq. (2)), i.e., $S_i^x = 2\operatorname{Re}(\rho_{10}^i)$, $S_i^y = 2\operatorname{Im}(\rho_{10}^i)$ and $S_i^z = 2\rho_{00}^i - 1$. Eqs. (6)-(8) can then be rewritten in the matrix form

$$\begin{pmatrix} S_i^x \\ S_i^y \\ S_i^z \end{pmatrix} = \sin^2(2a_i)\begin{pmatrix} P_i^x \\ P_i^y \\ P_i^z \end{pmatrix} + \cos(2a_i)\begin{pmatrix} \cos(2a_i) & -P_i^z \sin(2a_i) & P_i^y \sin(2a_i) \\ P_i^z \sin(2a_i) & \cos(2a_i) & -P_i^x \sin(2a_i) \\ -P_i^y \sin(2a_i) & P_i^x \sin(2a_i) & \cos(2a_i) \end{pmatrix}\begin{pmatrix} S_{i-1}^x \\ S_{i-1}^y \\ S_{i-1}^z \end{pmatrix}, (10)$$

or, equivalently, in the vector form

$$\vec{S}_i = \vec{P}_i \sin^2(2a_i) + \vec{S}_{i-1}\cos^2(2a_i) + \sin(2a_i)\cos(2a_i)\vec{P}_i \times \vec{S}_{i-1}. \quad (11)$$

This equation must be supplemented by the expressions for the density matrix at $t = 0$ and for the polarization vectors of the nuclear spins. Examples are considered below. All the results given in the figures below, were obtained, for definiteness, assuming that the initial electron spin polarization was 100%, in the positive direction along the $z$ axis, i.e., $S_0^x = S_0^y = 0$ and $S_0^z = 1$.

***Polarized nuclei.*** Let us assume an homogeneous nuclear spin polarization, that is $\vec{P}_i = \vec{P}$. It is convenient to introduce the transverse and longitudinal components of the electron spin polarization vector defined with respect to the direction of the nuclear spin polarization, $\vec{S}_i = \vec{S}_i^{\parallel} + \vec{S}_i^{\perp}$. The equations for the transverse and longitudinal polarization components can be easily decoupled,

$$\vec{S}_i^{\parallel} = \vec{P}_i \sin^2(2a_i) + \vec{S}_{i-1}^{\parallel} \cos^2(2a_i) \tag{12}$$

and

$$\vec{S}_i^{\perp} = \vec{S}_{i-1}^{\perp} \cos^2(2a_i) + \sin(2a_i)\cos(2a_i)\vec{P}_i \times \vec{S}_{i-1}^{\perp} \tag{13}$$

Consider Eq. (12). Expressing $\vec{S}_{i-1}^{\parallel}$ through $\vec{S}_{i-2}^{\parallel}$ and so on, we finally obtain

$$\vec{S}_i^{\parallel} = \vec{P}\left(\sin^2(2a_i) + \sin^2(2a_{i-1})\cos^2(2a_i) + ... + \sin^2(2a_1)\prod_{k=i}^{2}\cos^2(2a_k)\right) + \prod_{k=i}^{1}\cos^2(2a_k)\vec{S}_0^{\parallel} \tag{14}$$

After averaging over the ensemble of electrons, Eq. (14) takes the form

$$\vec{S}_i^{\parallel} = \vec{P}(1 - \cos^{2i}(2\bar{a})) + \cos^{2i}(2\bar{a})\vec{S}_0^{\parallel} = \vec{P} + \cos^{2i}(2\bar{a})(\vec{S}_0^{\parallel} - \vec{P}) = \vec{P} + e^{-\frac{t}{T_{\parallel}}}(\vec{S}_0^{\parallel} - \vec{P}) \quad, \tag{15}$$

where $\bar{a}$ is an average value of $a_i$, and the longitudinal spin relaxation time is introduced as

$$T_{\parallel} = -\frac{\Delta t}{2\ln(\cos(2\bar{a}))} \quad, \tag{16}$$

where $\Delta t$ is the average time between electron-nuclear spin interactions.

The dynamics of the transverse electron spin relaxation is governed by Eq. (13). The two terms on the right-hand side of Eq. (13) are perpendicular to each other. Evolution of the modulus of the transverse component of the electron spin $\vec{S}^{\perp}$, is given by

$$\left|\vec{S}_i^{\perp}\right| = \sqrt{\left|\vec{S}_{i-1}^{\perp}\right|^2 \cos^4(2a_i) + \sin^2(2a_i)\cos^2(2a_i)\left|\vec{P}_i\right|^2\left|\vec{S}_{i-1}^{\perp}\right|^2} = \left|\vec{S}_{i-1}^{\perp}\right|\cos(2a_i)\sqrt{\cos^2(2a_i) + \sin^2(2a_i)\left|\vec{P}_i\right|^2} \tag{17}$$

We can rewrite Eq. (17) in the form $\left|\vec{S}^{\perp}(t)\right| = \left|\vec{S}_0^{\perp}\right|e^{-\frac{t}{T_{\perp}}}$, with the transverse spin relaxation time

$$T_{\perp} = -\frac{\Delta t}{\ln\left(\cos(2\bar{a})\sqrt{\cos^2(2\bar{a}) + \sin^2(2\bar{a})\left|\vec{P}\right|^2}\right)} \quad. \tag{18}$$

Eq. (18) shows that the transverse spin relaxation time is a function of the nuclear spin polarization. Consider it in two limiting cases. If nuclear spins are unpolarized, $\left|\vec{P}\right| = 0$, then the transverse spin relaxation time is equal to the longitudinal spin relaxation time (Eq.(16)).

In the opposite limit, when the nuclear spins are 100% polarized, $|\vec{P}|=1$, the transverse relaxation time is longer by a factor of 2 then the longitudinal relaxation time. For intermediate values of the nuclear spin polarization, the transverse ratio is between these two limiting values.

Using Eq. (13), we can also calculate the angular velocity $\omega$ of the electron spin precession around the direction of the nuclear spin polarization. The angle between $\vec{S}_{i-1}$ and $\vec{S}_i$ is $\varphi = \arctan(|\vec{P}_i|\tan(2a_i))$. Since the direction of the electron spin polarization varies by $\varphi$ during the time interval $\Delta t$, the angular velocity is given by

$$\omega = \frac{\arctan(|\vec{P}_i|\tan(2a_i))}{\Delta t} \qquad . \qquad (19)$$

Using Eqs. (16), (18) for the relaxation times and Eq. (19) for the angular velocity, one finds the Bloch-type equation describing the evolution of the electron spin polarization

$$\frac{\partial \vec{S}}{\partial t} = \frac{\vec{P} - \vec{S}^{\parallel}}{T_{\parallel}} - \frac{\vec{S}^{\perp}}{T_{\perp}} + \omega \frac{\vec{P}}{|\vec{P}|} \times \vec{S} \qquad . \qquad (20)$$

Fig. 1 and Fig. 2 illustrate the results obtained in this part of the paper. Fig. 1 shows the longitudinal relaxation of the electron spin, when the nuclear spin polarization is defined as $P_i^x = P_i^y = 0$, $P_i^z = -1$ and the initial vector of the electron spin polarization is opposite to the nuclear spin polarization. The $z$-component of the electron spin polarization relaxes to the nuclear spin polarization with the relaxation time given by Eq. (16). The transverse relaxation of the electron spin polarization by the nuclear spins with the following polarization: $P_i^x = 1$, $P_i^y = P_i^z = 0$, is presented on Fig. 2. The oscillations of $\rho_{00}$ and $\text{Im}(\rho_{10})$ reflect the precession of the electron spin polarization vector around the direction of the nuclear spin polarization.

*Unpolarized nuclei.* It is important to study the electron spin relaxation by unpolarized nuclear spins. For zero polarization vector, $P_i^x = P_i^y = P_i^z = 0$, the evolution of the electron spin polarization is given by Eq. (20). The components of the electron spin polarization relax exponentially to the nuclear spin polarization with the relaxation time given by Eq. (16). This solution is shown on Fig. 3 by the dotted and dash lines (the initial electron spin polarization was selected again in the $+z$-direction).

Description of the nuclear spin polarization by the density matrix corresponds to the quantum-mechanical ensemble averaging over different system realizations. We have studied the noisy behavior of the electron spin relaxation when the system of Eqs. (6)-(8) is solved numerically assuming a single nuclear-spin-system realization, with random spin polarizations. The components of the nuclear spin polarization vector were selected randomly in the range $-1$ to $1$, with subsequent normalization of the polarization vector to 1. The results of our simulations are depicted on Fig. 3. It was found that for short times, the time dependence of the electron spin density matrix elements is sensitive to the realization. However for long times, the behavior of the diagonal matrix elements is close to exponential, while the off-diagonal matrix elements are close to zero, in agreement with the expected quantum-mechanically-averaged relaxation behavior.

*Discussion.* Let us estimate the electron spin relaxation time (Eq. (16)). Suppose that the electron wave packet is localized in the transverse directions on the length of the lattice constant $b$, and in the longitudinal direction on the de Broglie wavelength $\lambda$. The value of the electron wave function at a nucleus can be estimated as $|\psi(\vec{r}_i,t)|^2 = |u_0|^2 \left(\frac{\Omega}{b^2 \lambda}\right)$, where $\Omega$ is the unit cell volume, and the typical value of the square of the appropriately normalized Bloch function $|u_0|^2$ for semiconductors is $|u_0|^2 = 5 \cdot 10^{25}$ cm$^{-3}$ [28]. For the density of one nuclear spin-1/2 per unit cell, we obtain $\Delta t = \frac{b}{v}$, where the average electron velocity can be taken $v = 10^7$ cm/sec, and, typically, $b = 5$ Å. Since the longitudinal electron wave packet size is of order $\lambda$, we can define the effective electron-nuclear spin interaction time as $\delta_i t = \frac{\lambda}{v}$. The final results do not depend on the Broglie wavelength $\lambda$ since the constant $\bar{a}$ is proportional to the product of the wave function square on the nucleus and the effective electron-nuclear spin interaction time. With these assumptions, our estimation of the electron spin relaxation time is $\tau_0 = 30$ nsec.

To conclude, we have studied the electron spin relaxation by nuclear spins in semiconductors in the framework of a semi-classical approach to the spatial motion. We have obtained equations describing the evolution of the electron spin density matrix elements and have solved them in illustrative cases, namely for the cases of the fully polarized nuclei and unpolarized nuclei. Under conditions assumed, the electron spin polarization relaxes in time to the nuclear spin polarization. The observed effect can be used to create spin-polarized electrons in spintronic devices. The electron spin relaxation times were estimated.

We gratefully acknowledge helpful discussions with Professor I. D. Vagner. This research was supported by the National Science Foundation, grants DMR-0121146 and ECS-0102500, and by the National Security Agency and Advanced Research and Development Activity under Army Research Office contract DAAD 19-02-1-0035.

**References.**

[1] S. A. Wolf, D. D. Awschalom, R. A. Buhrman, J. M. Daughton, S. von Molnár, M. L. Roukes, A. Y. Chtchelkanova and D. M. Treger, Science **294**, 1488 (2001).

[2] S. Datta and B. Das, Appl. Phys. Lett. **56**, 665 (1990).

[3] J. Nitta, T. Akazaki, H. Takayanagi and T. Enoki, Phys. Rev. Lett. **78**, 1336 (1997).

[4] G. Meir, T. Matsuyama and U. Merkt, Phys. Rev. B **65**, 125327 (2002).

[5] J. Fabian, I. Žutić and S. Das Sarma, Phys. Rev. B **66**, 165301 (2002).

[6] P. Mohanty, Physica B **280**, 446 (2000).

[7] *Optical Orientation*, edited by B. Meier and B. P. Zakharchenya (North Holland, Amsterdam, 1984), Ch. III.

[8] P. H. Song and K. W. Kim, Phys. Rev. B **66**, 035207 (2002).

[9] G. Salis, D. T. Fuchs, J. M. Kikkawa, D. D. Awschalom, Y. Ohno and H. Ohno Phys. Rev. Lett. **86**, 2677 (2001).

[10] A. V. Khaetski, D. Loss and L. Glazman, Phys. Rev. Lett. **88**, 186802 (2002).

[11] I. A. Merkulov, Al. L. Efros and M. Rosen, Phys. Rev. B **65**, 205309 (2002).

[12] S. I. Erlingsson, Y. L. Nazarov and V. Fal'ko, Phys. Rev. B **64**, 195306 (2001).

[13] R. de Sousa and S. Das Sarma, preprint cond-mat/0211567 (2002).

[14] S. Saykin, D. Mozyrsky and V. Privman, Nano Letters **2**, 651 (2002).

[15] R. J. Elliot, Phys. Rev. **96**, 266 (1954).

[16] Y. Yafet, in *Solid State Physics*, edited by F. Seitz and D. Turnbull (Academic, New York, 1963), Vol. 14.

[17] M. I. D'yakonov and V. I. Perel, Zh. Éksp. Teor. Fiz. **60**, 1954 (1971) [Sov. Phys. JETP **33**, 1053 (1971)]; Fiz. Tverd. Tela (Leningrad) **13**, 3581 (1971) [Sov. Phys. Solid State **13**, 3023 (1972)].

[18] G. L. Bir, A. G. Aronov and G. E. Pikus, Zh. Éksp. Teor. Fiz. **69**, 1382 (1975) [Sov. Phys. JETP **42**, 705 (1976)].

[19] Y. Ohno, R. Terauchi, T. Adachi, F. Matsukura and H. Ohno, Phys. Rev. Lett. **83**, 4196 (1999).

[20] B. E. Kane, L. N. Pfeiffer, and K. W. West, Phys. Rev. B **46**, 7264 (1992).

[21] K. R. Wald, L. P. Kouwenhoven, P. L. McEuen, N. C. van der Vaart, and C. T. Foxon, Phys. Rev. Lett. **73**, 1011 (1994).

[22] D. C. Dixon, K. R. Wald, P. L. McEuen, and M. R. Melloch, Phys. Rev. B **56**, 4743 (1997).

[23] T. Machida, T. Yamazaki, and S. Komiyama, Appl. Phys. Lett. **80**, 4178 (2002).

[24] A. M. Dyugaev, I. D. Vagner and P. Wyder, JETP Lett. **64**, 207 (1996) ;cond-mat/0005005.

[25] R. K. Kawakami, Y. Kato, M. Hanson, I. Malajovich, J. M. Stephens, E. Johnston-Halperin, G. Salis, A.C. Gossard, and D. D. Awschalom, Science **294**, 131 (2001).

[26] J. H. Smet, R. A. Deutschmann, F. Ertl, W. Wegscheider, G. Abstreiter, and K. von Klitzing, Nature **415**, 281 (2002).

[27] S. Rehmann, T. Herrmannsdörfer, and F. Pobell, Phys. Rev. Lett. **78**, 1122 (1997).

[28] D. Paget, G. Lampel, B. Sapoval and V. I. Safarov, Phys. Rev. B **15**, 5780 (1977).

[29] Yu. V. Pershin, S. N. Shevchenko, I. D. Vagner, and P. Wyder, Phys. Rev. B **66**, 035303 (2002).

[30] C. P. Slichter, *Principles of Magnetic Resonance* (Springer-Verlag, Berlin, 1991), 2nd ed.

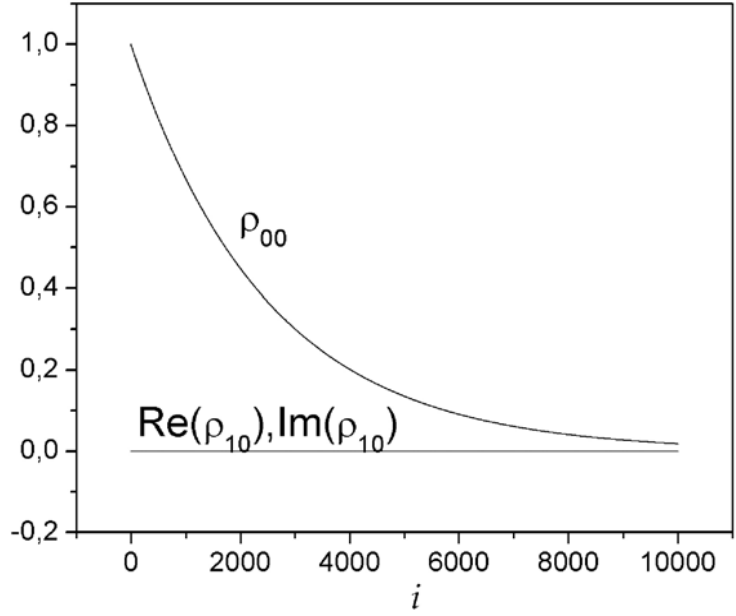

**Fig. 1.** Evolution of the electron spin density matrix due to the interaction with the nuclear spins completely polarized in the (–z)-direction, $P_i^x = P_i^y = 0$, $P_i^z = -1$, $a_i = 0.01$.

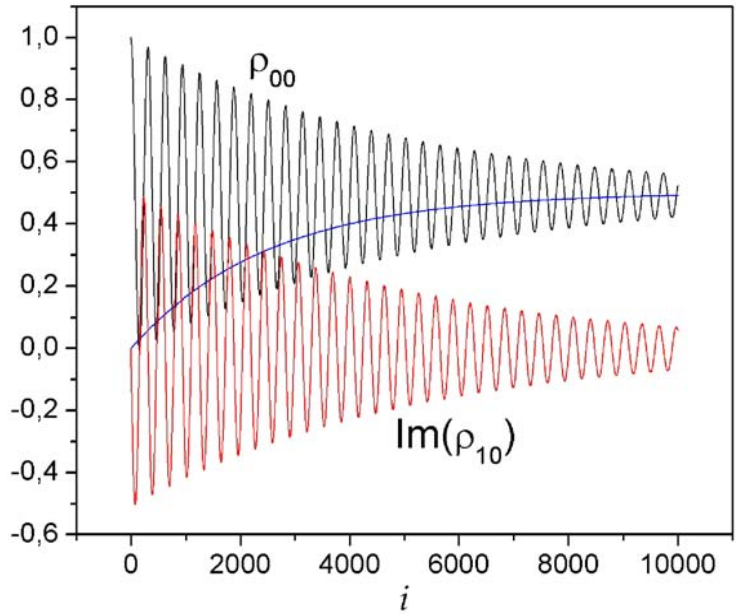

**Fig. 2.** Evolution of the electron spin density matrix due to the interaction with completely polarized nuclear spins in the +x-direction, $P_i^x = 1$, $P_i^y = P_i^z = 0$, $a_i = 0.01$. The blue line is $\text{Re}(\rho_{10})$.

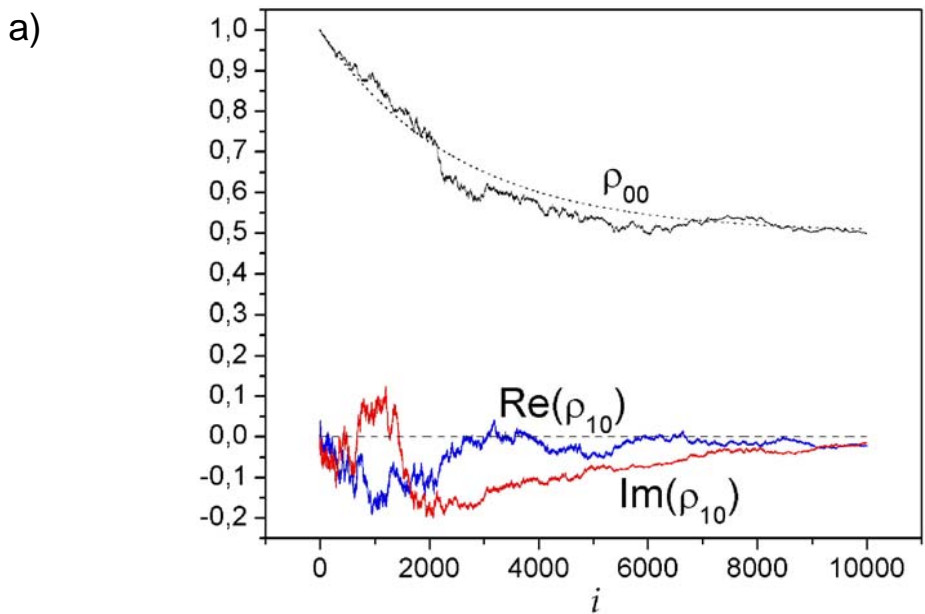

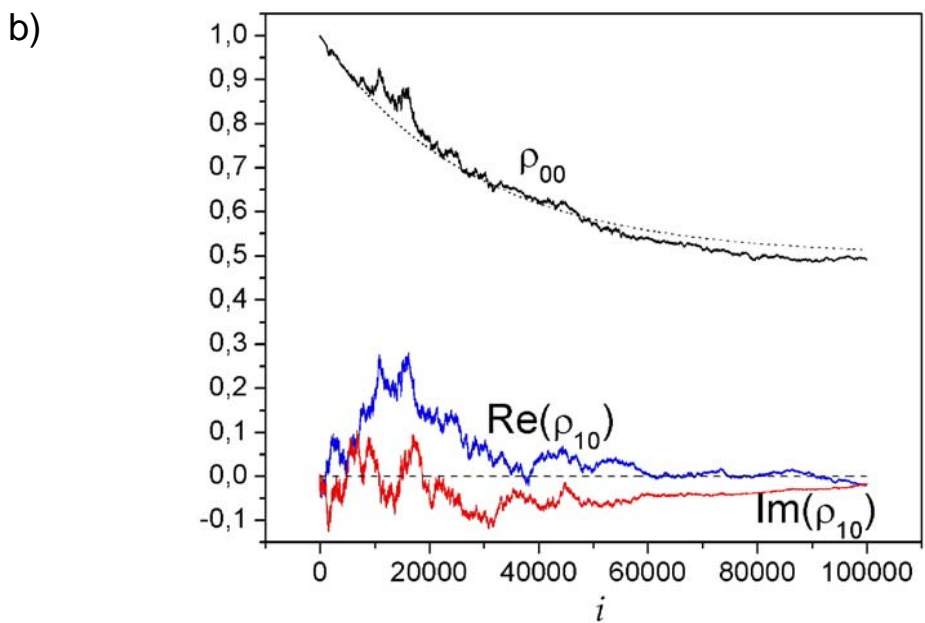

**Fig. 3.** Evolution of the electron spin density matrix elements caused by the interaction with unpolarized nuclear spins. Unpolarized nuclear spins were modeled by zero polarization vector (dotted and dashed lines) and by unit polarization vector directed randomly (noisy data lines): a) $a_i = 0.01$; b) $a_i = 0.003$.